\documentclass[prc,aps,twocolumn,showpacs,superscriptaddress,floatfix]{revtex4}
\def\FigFactor{0.5}

\usepackage{graphics}
\def\pip {$\pi^+$}
\def\pim {$\pi^-$}

\def\dedx {$dE/dx$}
\def\meandedx {$\langle dE/dx \rangle$}

\def\mtm {$m_t-m_0$}
\def\mt {$m_t$}

\begin{document}

\title{Charged Pion Production in 2 to 8 AGeV Central Au+Au 
Collisions}

\author{J.L.~Klay}      \affiliation{University of California, Davis, California 95616}
\author{N.N.~Ajitanand} \affiliation{Department of Chemistry and Physics, SUNY, Stony Brook, New York 11794-3400}
\author{J.M.~Alexander} \affiliation{Department of Chemistry and Physics, SUNY, Stony Brook, New York 11794-3400}
\author{M.G.~Anderson}  \affiliation{University of California, Davis, California 95616}
\author{D.~Best}        \affiliation{Lawrence Berkeley National Laboratory, Berkeley, California 94720}
\author{F.P.~Brady}     \affiliation{University of California, Davis, California 95616}
\author{T.~Case}        \affiliation{Lawrence Berkeley National Laboratory, Berkeley, California 94720}
\author{W.~Caskey}      \affiliation{University of California, Davis, California 95616}
\author{D.~Cebra}       \affiliation{University of California, Davis, California 95616}
\author{J.L.~Chance}    \affiliation{University of California, Davis, California 95616}
\author{P.~Chung}       \affiliation{Department of Chemistry and Physics, SUNY, Stony Brook, New York 11794-3400}
\author{B.~Cole}        \affiliation{Columbia University, New York, New York 10027}
\author{K.~Crowe}       \affiliation{Lawrence Berkeley National Laboratory, Berkeley, California 94720}
\author{A.C.~Das}       \affiliation{The Ohio State University, Columbus, Ohio 43210}
\author{J.E.~Draper}    \affiliation{University of California, Davis, California 95616}
\author{M.L.~Gilkes}    \affiliation{Purdue University, West Lafayette, Indiana 47907-1396}
\author{S.~Gushue}      \affiliation{Brookhaven National Laboratory, Upton, New York 11973}
\author{M.~Heffner}     \affiliation{University of California, Davis, California 95616}
\author{A.S.~Hirsch}    \affiliation{Purdue University, West Lafayette, Indiana 47907-1396}
\author{E.L.~Hjort}     \affiliation{Purdue University, West Lafayette, Indiana 47907-1396}
\author{L.~Huo}         \affiliation{Harbin Institute of Technology, Harbin, 150001 People's Republic of China}
\author{M.~Justice}     \affiliation{Kent State University, Kent, Ohio 44242}
\author{M.~Kaplan}      \affiliation{Carnegie Mellon University, Pittsburgh, Pennsylvania 15213}
\author{D.~Keane}       \affiliation{Kent State University, Kent, Ohio 44242}
\author{J.C.~Kintner}   \affiliation{St. Mary's College of California, Moraga, California 94575}
\author{D.~Krofcheck}   \affiliation{University of Auckland, Auckland, New Zealand}
\author{R.A.~Lacey}     \affiliation{Department of Chemistry and Physics, SUNY, Stony Brook, New York 11794-3400}
\author{J.~Lauret}      \affiliation{Department of Chemistry and Physics, SUNY, Stony Brook, New York 11794-3400}
\author{C.~Law}         \affiliation{Department of Chemistry and Physics, SUNY, Stony Brook, New York 11794-3400}
\author{M.A.~Lisa}      \affiliation{The Ohio State University, Columbus, Ohio 43210}
\author{H.~Liu}         \affiliation{Kent State University, Kent, Ohio 44242}
\author{Y.M.~Liu}       \affiliation{Harbin Institute of Technology, Harbin, 150001 People's Republic of China}
\author{R.~McGrath}     \affiliation{Department of Chemistry and Physics, SUNY, Stony Brook, New York 11794-3400}
\author{Z.~Milosevich}  \affiliation{Carnegie Mellon University, Pittsburgh, Pennsylvania 15213}
\author{G.~Odyniec}     \affiliation{Lawrence Berkeley National Laboratory, Berkeley, California 94720}
\author{D.L.~Olson}     \affiliation{Lawrence Berkeley National Laboratory, Berkeley, California 94720}
\author{S.Y.~Panitkin}  \affiliation{Kent State University, Kent, Ohio 44242}
\author{C.~Pinkenburg}  \affiliation{Department of Chemistry and Physics, SUNY, Stony Brook, New York 11794-3400}
\author{N.T.~Porile}    \affiliation{Purdue University, West Lafayette, Indiana 47907-1396}
\author{G.~Rai}         \affiliation{Lawrence Berkeley National Laboratory, Berkeley, California 94720}
\author{H.G.~Ritter}    \affiliation{Lawrence Berkeley National Laboratory, Berkeley, California 94720}
\author{J.L.~Romero}    \affiliation{University of California, Davis, California 95616}
\author{R.~Scharenberg} \affiliation{Purdue University, West Lafayette, Indiana 47907-1396}
\author{B.~Srivastava}  \affiliation{Purdue University, West Lafayette, Indiana 47907-1396}
\author{N.T.B.~Stone}   \affiliation{Lawrence Berkeley National Laboratory, Berkeley, California 94720}
\author{T.J.M.~Symons}  \affiliation{Lawrence Berkeley National Laboratory, Berkeley, California 94720}
\author{S.~Wang}        \affiliation{Kent State University, Kent, Ohio 44242}
\author{R.~Wells}       \affiliation{The Ohio State University, Columbus, Ohio 43210}
\author{J.~Whitfield}   \affiliation{Carnegie Mellon University, Pittsburgh, Pennsylvania 15213}
\author{T.~Wienold}     \affiliation{Lawrence Berkeley National Laboratory, Berkeley, California 94720}
\author{R.~Witt}        \affiliation{Kent State University, Kent, Ohio 44242}
\author{L.~Wood}        \affiliation{University of California, Davis, California 95616}
\author{W.N.~Zhang}     \affiliation{Harbin Institute of Technology, Harbin, 150001 People's Republic of China}
\collaboration{E895 Collaboration}      \noaffiliation

\date{\today}   

\begin{abstract}
Momentum spectra of charged pions over nearly full rapidity coverage 
from target to beam rapidity have been extracted from 0-5\% most central 
Au+Au collisions in the beam energy range from 2 to 8 AGeV by the E895 Experiment.  Using a 
thermal parameterization to fit the transverse mass spectra, rapidity 
density distributions are extracted.  The observed spectra are compared 
with predictions from the RQMD v2.3 cascade model and also to a thermal 
model including longitudinal flow.  The total 4$\pi$ yields of the charged 
pions are used to infer an initial state entropy produced in the collisions.
\end{abstract}


\maketitle

\section{Introduction}

One of the primary goals of the study of heavy ion collisions at relativistic
energies is the improvement of our understanding of the bulk properties
of nuclear/hadronic matter at high temperatures and densities.  As a first
step in understanding these properties, one should carefully characterize
the particle species that make up the bulk of the matter.  For the energy
regime of the Bevalac/SIS (0.2 to 1.2 AGeV), collisions between heavy
nuclei cause a compression of the nuclear matter, resulting in a
disassembly into the constituent neutrons and protons, which are emitted
either individually or bound within various light composite fragments 
(d, t, $^{3}$He, $^{4}$He)\cite{Lisa95,Herr96}.  

At the top energy of AGS Au+Au collisions of 10.8 AGeV, the most copiously produced 
charged particles are the lightest of the mesons, the pions\cite{Ahle98,Barr00}. 
The matter has evolved from a heated and compressed gas of nucleons into 
a hot dense gas of hadrons, predominantly pi mesons.  Thus the energy 
region from 2 to 8 AGeV represents a transition regime.  Studies of nuclear stopping 
suggest the maximum density achieved increases from three times normal nuclear density 
at 1 AGeV\cite{Stocker86,Bert88,Wang91} to eight times normal nuclear density at 10 
AGeV\cite{LiKo95}.  Measurements of the proton directed and elliptic
flow indicate that hydrodynamic flow has saturated across this energy
range and cannot account for the increased energy available\cite{Pink99,Liu00}. 
This additional available energy goes primarily into pion production.  By studying the 
pion production across this regime we are able to observe nuclear matter in transition.

Early on in the study of heavy ion collisions, an enhanced yield of pions was seen as a 
possible signature of a transition to a deconfined state of matter.  However, in the 
early studies of pion yields in the 1 AGeV energy range at the 
Bevalac\cite{Naga81,Strob83,Stoc86}, it was observed that the measured pion production 
cross sections were smaller than predicted at the time. This observation led to the
conjecture, which was later experimentally demonstrated, that the excess
kinetic energy was converted into hydrodynamical flow effects. The strong
radial flow observed at this energy implies a significant expansion and
cooling, which limits the freeze-out pion multiplicities\cite{Dani95,Stocker84}.
More recently, more detailed mesurements of the pion yields from 1 AGeV Au+Au
collisions have become available from the SIS 
experiments\cite{Pelt97,Wagn98,Vene93,Schwalb94}.  These results demonstrate a roughly 
two to one ratio of $\pi^-$ over $\pi^+$ and a strong non-thermal low-p$_t$ 
enhancement.  Both features are strong indicators that pions at this energy are 
produced almost exclusively through the Delta resonance\cite{Sorge94}. 
In full energy AGS collisions (Au+Au at 10.8 AgeV), pion production has been 
studied at midrapidity\cite{Ahle98} and at target rapidity\cite{Barr95}.
Although there is still an asymmetry between $\pi^-$ and $\pi^+$ production 
and there is also still evidence of a low-p$_t$ enhancement in the momentum spectra, 
these effects are much less significant than at SIS.  The broad rapidity coverage for 
both pions and protons at these energies has also been used to study the development of 
longitudinal flow\cite{Stac96}.

This paper will detail the development of pion production across the beam
energy range from 2 to 8 AGeV.  The role of the Delta resonance production
mechanism will be explored through observations of the overall pion 
ratios and through RQMD simulations.  The rapidity density distributions 
will be used with previously published proton rapidity distributions\cite{Klay02} to 
explore the effects of collective longitudinal flow.  The overall pion 
multiplicities will be used to establish a low energy baseline for a recent pion 
multiplicity based QGP search at the CERN SPS\cite{NA4902}.

\section{Data Collection and Analysis}
        The data were taken at the Brookhaven National Laboratory
Alternating Gradient Synchrotron (AGS) by the E895 Experiment using the
EOS TPC \cite{EOSTPC} during a series of runs in 1996.  This article
presents charged pion transverse mass and rapidity density spectra from
Au+Au collisions at nominal beam energies of 2, 4, 6, and 8 GeV/nucleon
(AGeV). (After correcting for energy loss before the target, the actual
beam energies of collisions at 2 and 4 AGeV were found to be 1.85 and 3.91
AGeV, respectively.  No corrections were necessary for 6 and 8 AGeV.)  
The EOS TPC provides nearly 4$\pi$ solid angle coverage, which makes
global characterization of the collision events possible.  Charged 
particle momenta are reconstructed from the helical trajectories of tracks 
reconstructed from the ionization trails left by particles passing 
through the TPC, which was situated inside the Multi-Particle 
Spectrometer (MPS) magnet.  Data presented for 2 AGeV collisions were 
taken in a 0.75 Tesla field, while 4, 6, and 8 AGeV collisions were taken 
in a 1 Tesla field.  

	A primary track multiplicity for each event is obtained 
by rejecting those tracks which do not pass within 2.5 cm of the 
reconstructed event vertex in the target.  The multiplicity 
distribution from a minimum trigger-biased sample of events is 
used to discriminate event centrality classes by assuming a monotonic 
relationship between the impact parameter and multiplicity \cite{Cava90}.  
The data presented in this paper are selected on the top 5\% most central 
events; an ensemble of approximately 20,000 events at each energy 
for this centrality selection were used to obtain the spectra.

Identification of particle species is determined via multiple sampling 
(up to 128 samples) of the ionization in P10 (10\% methane, 
90\% argon) drift gas.  The average ionization energy loss, {\meandedx}, 
is computed for each track from the available samples via a truncated 
mean to reduce the influence of the large energy tail of the 
distribution.   The 20\% highest {\dedx} samples are discarded from the 
calculation of the average.  

\begin{figure}
\resizebox*{\FigFactor\textwidth}{!}{\includegraphics{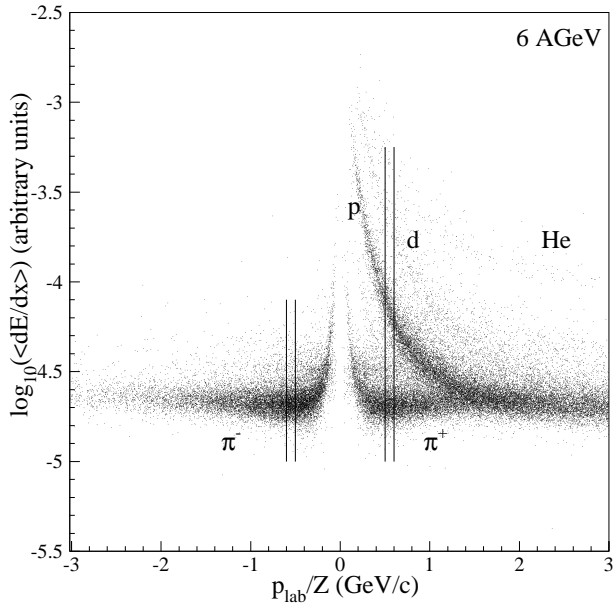}}
\caption{{\meandedx} as a function of rigidity at 6 AGeV.  A 
parameterized Bethe-Bloch prediction is used to fix the mean
energy loss for each particle species as a function of rigidity.  The narrow slices
illustrate the location of the {\meandedx} projections shown in Fig.~\ref{fig:slice_example}.}
\label{fig:dedx_map}
\end{figure}

Fig.~\ref{fig:dedx_map} shows a typical particle identification map 
with {\meandedx} plotted versus reconstructed track rigidity (= p/Z) at 
6 AGeV.  The various particle species are identified by their separation 
into bands.  The pion spectra were obtained by fitting projections of the {\meandedx} 
in narrow bins of {\mtm} and rapidity, using an assumption of the pion mass 
and charge to calculate the ({\mtm},y) coordinates for a given measured 
particle ($p_x$,$p_y$,$p_z$).  

The single particle {\meandedx} projections are often described by a Gaussian 
distribution centered on the mean value predicted by a Bethe-Bloch formulation.  
However, this assumes that the calculation of {\meandedx} for each track was obtained 
in an identical fashion for all tracks in the distribution.  In fact, the 
truncated mean method used in this analysis introduces a skewing toward 
larger {\meandedx} which comes from combining tracks of different number 
of samples, N$_{hits}$.  One way to avoid this skewing is to divide the 
data into bins of N$_{hits}$, which reduces the effect.  However the 
reduction in statistics for each bin leads to increased uncertainty in 
the determined yields.  Therefore, for this analysis, a model of the 
N$_{hits}$-integrated single particle {\meandedx} distribution shapes 
was used with the predicted mean values from a Bethe-Bloch parameterization of the 
{\meandedx} as a function of $\beta\gamma$ to extract the total yields of pions from 
each projection.  The model is represented by a correlated sum of Gaussian 
distributions:  a main Gaussian for the bulk of the distribution plus a smaller, 
offset ``shoulder'' Gaussian for the high-{\meandedx} tail.  The parameters of the 
model were studied as a function of beam energy, particle type and ({\mtm},y) bin in 
regions where the single particle distributions can be separately characterized.  
Tight constraints on the model parameters were applied in order to 
extrapolate the model into regions where multiple particle distributions 
partially overlap.  Thus the relative yields of different particles were 
deconvoluted in each ({\mtm},y) bin:  0.1 unit rapidity slices over the 
full rapidity range from target to beam rapidity, with the midrapidity 
bin covering the range -0.05 $<$ y$_{cm}$ $<$ 0.05, and in 25 MeV/c$^{2}$ 
bins in m$_{t}$-m$_{0}$ in the range 0 $<$ m$_{t}$-m$_{0}$ $<$ 1.0 
GeV/c$^{2}$.  Particles of the wrong mass and/or charge contaminating the 
pion sample in a given ({\mtm},y) bin were discarded.

\begin{figure}
\resizebox*{\FigFactor\textwidth}{!}{\includegraphics{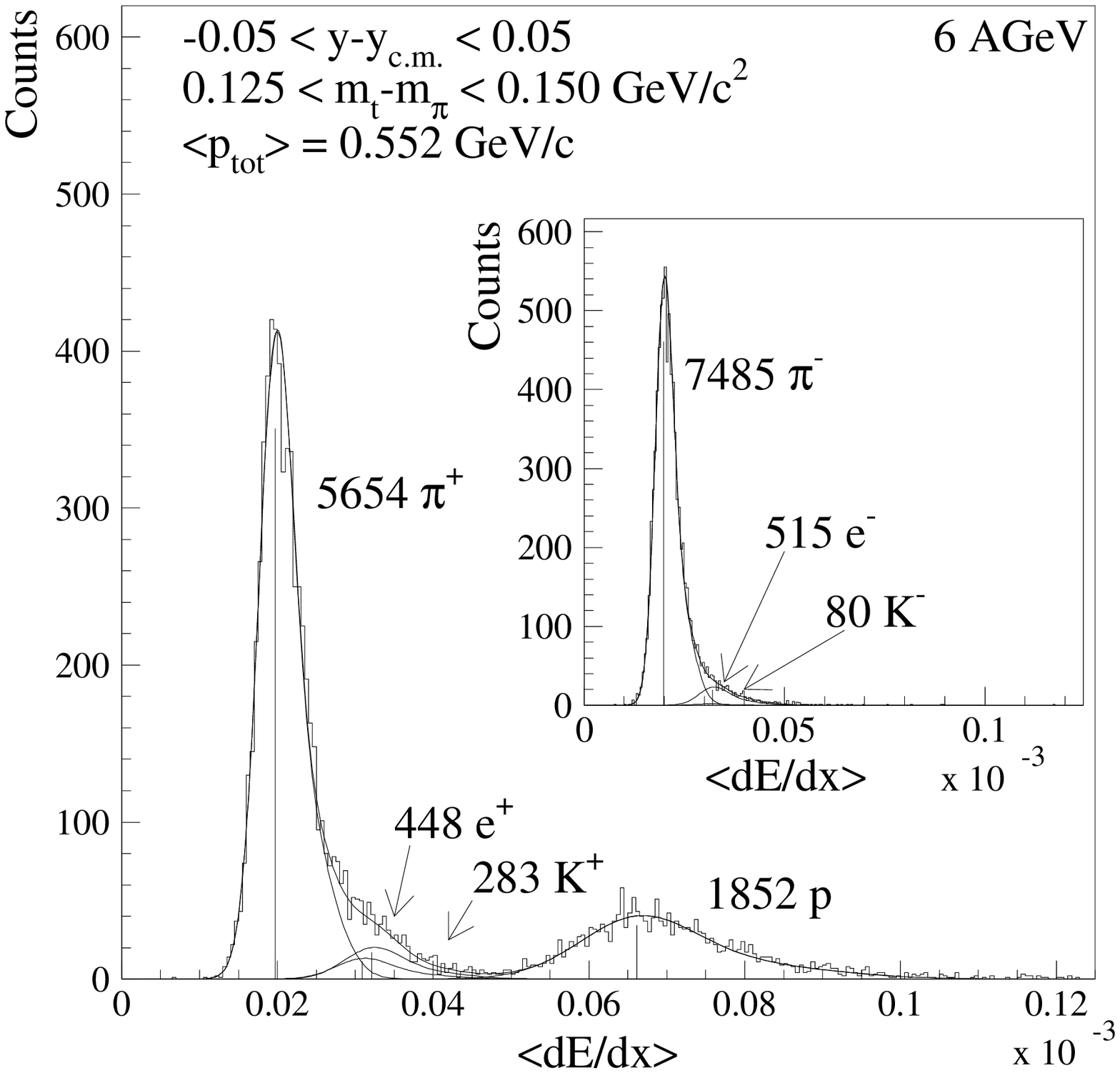}}
\caption{Projections of the positive and negative (inset) particle {\meandedx} at mid-rapidity and 0.125 $<$ {\mtm} $<$ 
0.150 Gev/c$^2$, in the rigidity ranges indicated on Fig.~\ref{fig:dedx_map}.  The ({\mtm},y) ranges were calculated 
assuming the pion mass.  The function used to fit the data is the two-Gaussian model 
(described in the text) of the single particle {\meandedx} distributions, one for each particle type.  The integrals under the pion peaks 
are the total yield for the fitted ({\mtm},y) bin and the contributions from all other particles are discarded.}
\label{fig:slice_example}
\end{figure}

Fig.~\ref{fig:slice_example} shows an example of a single ({\mtm},y) slice at 
midrapidity for 6 AGeV and $0.125<${\mtm}$<0.150$ GeV/c$^2$.  The total yields 
are obtained by integrating the area under the fitted distributions for each 
particle.  Oppositely charged particles are projected separately.  The inset shows 
the projection of negatively charged particles, which are well-characterized by this 
method of particle identification over the entire range of transverse mass and 
rapidity.  However, as the rapidity and transverse mass increase, the positively 
charged pions suffer increasing contamination due to the overlap with the positive 
kaons and protons.  The range of {\mtm} and rapidity over which the $\pi^+$ yields 
are extracted is therefore more limited.

Electron contamination was estimated by studying their yields in the regions of phase 
space where they are relatively cleanly identified (for p$_{lab}$ $<$ 150 MeV/c and 
300 $<$ p$_{lab}$ $<$ 500 MeV/c) and interpolating between these limits.  It was found 
that electrons contribute approximately 10\% to the observed yield of pions for lab 
momenta 150 $<$ p$_{lab}$ $<$ 250 MeV/c.  Since the electron yields fall significantly 
as a function of their own transverse momentum, the contamination predominantly 
effects the lowest pion {\mtm} bins.  The errors on the quoted pion yields account for 
possible systematic uncertainties in the determination of the electron contamination. 

Observed kaon yields from the same beam energy range, measured by E866/E917 
\cite{Ogil98,Dunl99}, folded with the EOS TPC detector response allows us to extend 
the reach of the {\pip} fitting out to p$_{lab}$ $\sim$ 1.2 GeV/c.  
For lab momenta above 1.2 GeV/c, however, the $\pi^+$ become hopelessly 
entangled with the protons, which are approaching minimum ionizing {\meandedx}.  This 
cut-off manifests at different {\mtm}, depending on the rapidity and bombarding energy.

Detector response to the final-state collision products has been extensively studied 
using the GEANT 3.21 simulation package.  Small (maximum of 4 per event) samples of 
pions with momentum distributions approximating the real data are embedded into full 
data events.  These particles are tagged and propagated through the data 
reconstruction chain to determine the effects of detector acceptance, tracking 
efficiency, and momentum resolution.  

\begin{figure}
\resizebox*{\FigFactor\textwidth}{!}{\includegraphics{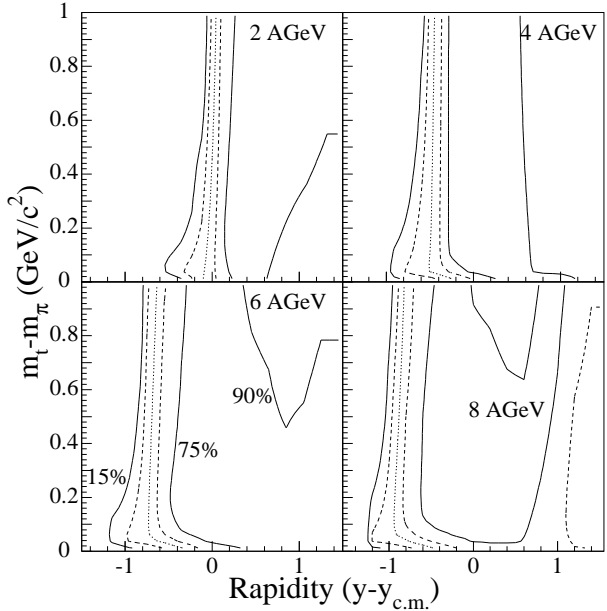}}
\caption{EOS TPC total ``efficiency'' ($\equiv$ 1/correction, including effects of 
acceptance, tracking efficiency and momentum resolution) for pions produced in 2, 4, 
6 and 8 AGeV 0-5\% central Au+Au collisions.  The efficiencies are obtained from 
GEANT simulations of the detector response to Monte Carlo tracks embedded into 
real data events and propagated through the reconstruction chain.  The contours, 
shown in step of $\sim$15\%, include the effects of geometric acceptance, tracking 
efficiency and momentum resolution.}
\label{fig:effic}
\end{figure}

Since the beam actually passes through the sensitive volume of the detector, there is 
no explicit low-p$_t$ measurement cut-off.  However, forward-focusing causes increased 
tracking losses at low p$_t$ and forward rapidities due to higher track densities 
and track merging.  Losses at backward rapidities and high {\mtm} are dominated by the 
geometric acceptance of the detector, while at more forward rapidities, the high 
{\mtm} tracks are stiffer and therefore suffer from worsening momentum resolution.   
Fig.~\ref{fig:effic} shows the overall detection efficiency obtained from 
simulations as a function of {\mtm} and rapidity for pions at each beam energy.   The 
contours indicate steps of $\sim$15\%.   The raw data are corrected for these effects 
to extract the total yields of positive and negative pions as a function of {\mtm} and 
rapidity at each beam energy.  

Due to the large acceptance of our device, we can test the systematic uncertainties on the yields by 
comparing forward and backward rapidity corrected spectra.  Overlapping acceptances 
at 2, 4, 6 and 8 GeV allow us to conclude that rapidity bins corresponding to 
$y_{lab}<$ 0.5 are most affected by the systematic 
uncertainties in our corrections.  The underestimated corrections in this region of phase 
space are not surprising given that neither detector performance nor detector simulations were optimized 
for target rapidity in the lab frame.  Here tracks cross the fewest pad rows and have the smallest radii 
of curvature.  Mid-rapidity yields at all four beam energies were checked against published results 
from E866/E917\cite{Seto98,Ahle00}.  The average level of agreement between our spectra and E866/E917 over 
all {\mtm} and beam energies is observed to be approximately 7\%.  The minimum systematic uncertainties 
at all rapidities for the E895 data presented here are estimated to be 5\%, while 
for the most backward rapidity bins, the uncertainties at high {\mtm} can become as 
large as 50\%.  These errors are included on the spectra reported in this 
article.

\section{Results}

\begin{figure*}[th]
\resizebox*{0.7\textheight}{!}{\includegraphics{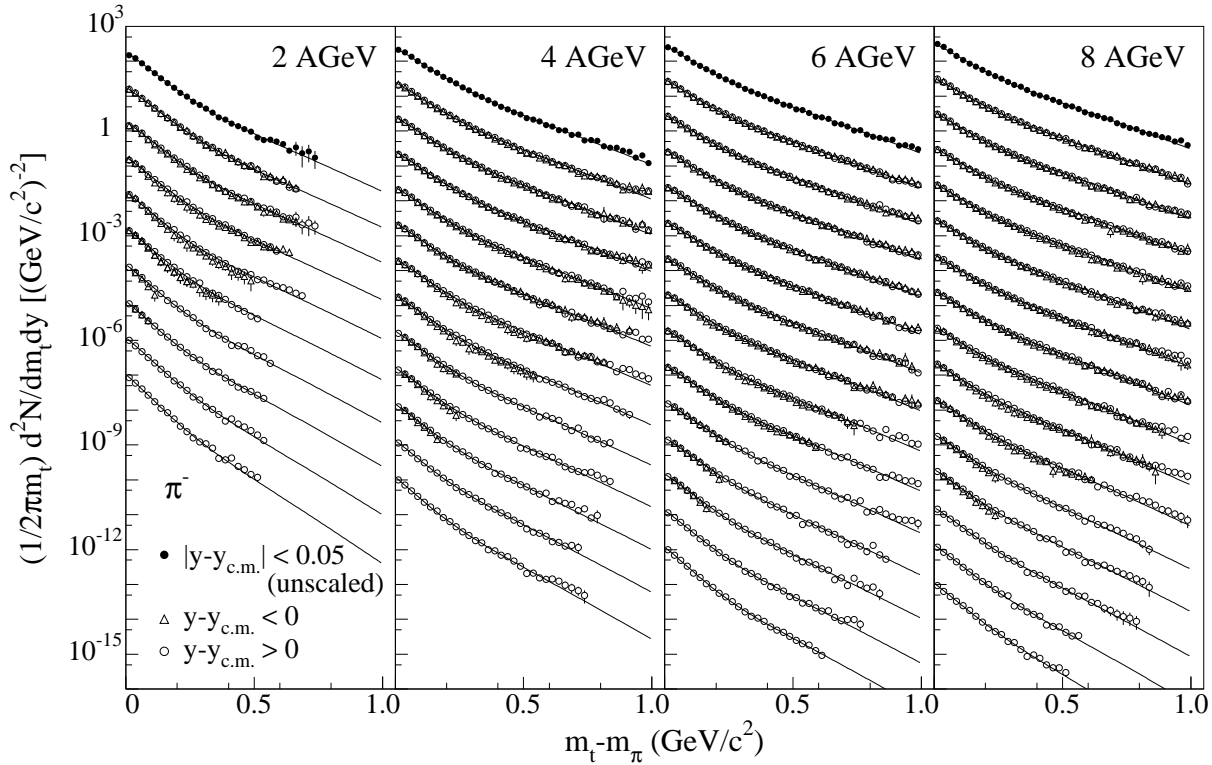}}
\caption{Invariant yield per event as a function of {\mtm} for {\pim} at 2, 4, 6, and 8 AGeV.
Midrapidity is shown unscaled, while the 0.1 unit forward/backward rapidity slices are scaled down
by successive factors of 10.  The functions plotted with the data are two-slope 
Boltzmann parameterizations described in the text.}
\label{fig:pim_mt}
\end{figure*}

\begin{figure*}[th]
\resizebox*{0.7\textheight}{!}{\includegraphics{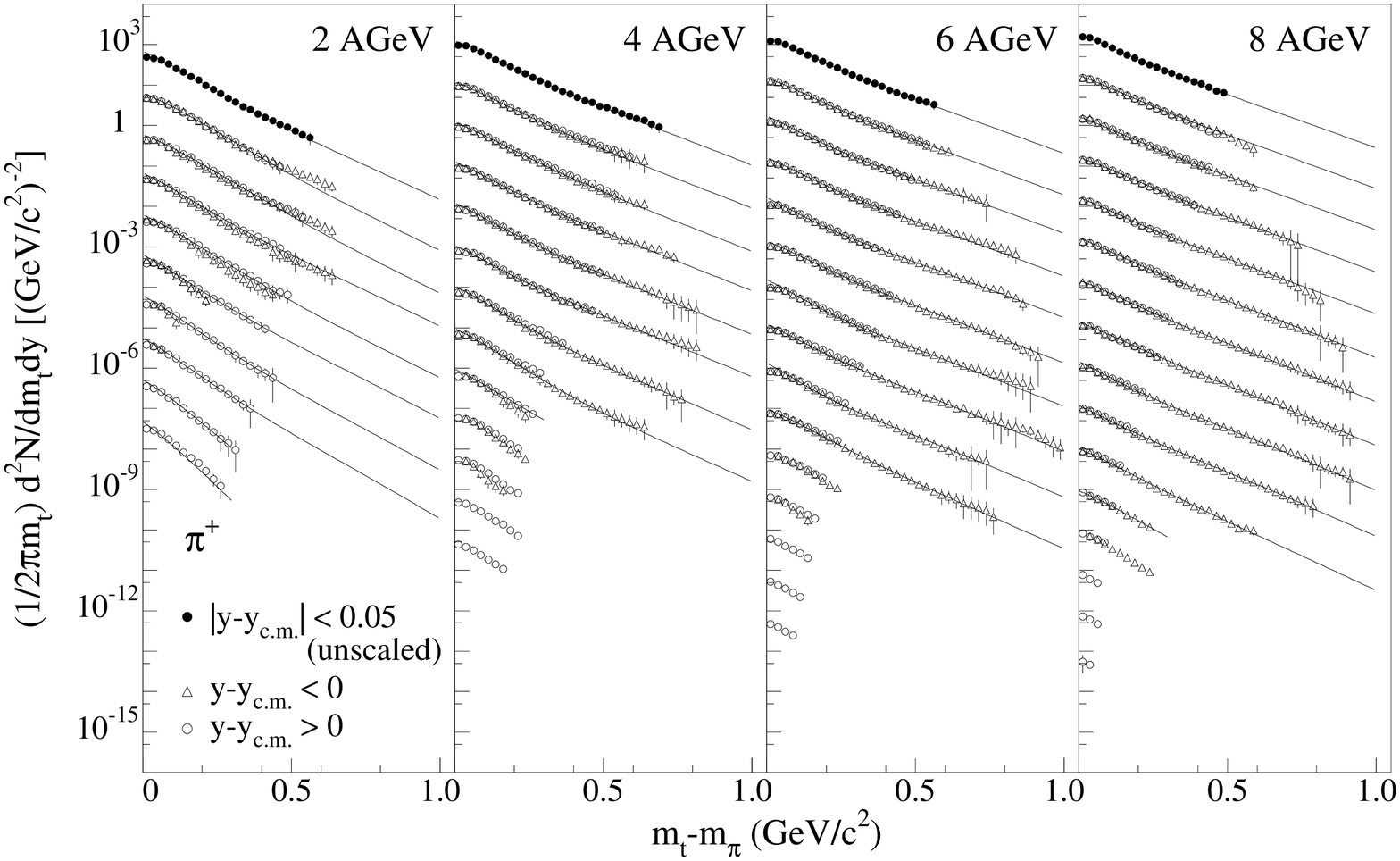}}
\caption{Invariant yield per event as a function of {\mtm} for {\pip} at 2, 4, 6, and 8 AGeV.
Midrapidity is shown unscaled, while the 0.1 unit forward/backward rapidity slices are scaled down
by successive factors of 10. The functions plotted with the data are two-slope 
Boltzmann parameterizations described in the text.}
\label{fig:pip_mt}
\end{figure*}

	Figs.\ \ref{fig:pim_mt} and \ref{fig:pip_mt} show the fully 
corrected, invariant yields of charged pions per event from 0-5\% central Au+Au 
collisions at 2,4,6 and 8 AGeV.  Mid-rapidity is shown unscaled as black 
circles, while each bin forward/backward of mid-rapidity is scaled down by 
a successive factor of 10.  Forward rapidities are indicated as open 
circles and backward rapidities as open triangles. The reported error bars include both statistical 
and systematic uncertainties.

The approximately exponential decay of the particle yields as a function of 
transverse mass has been observed in high energy particle and heavy ion experiments over a wide 
range of conditions.  In order to extract the full 4$\pi$ yields of pions (including 
those from resonance feed-down) produced in the beam energy range studied here, a 
simple parameterization of the pion {\mtm} spectra which reproduces the observed 
shapes of the spectra over the full range of {\mtm}, from 0 to 1 GeV/c$^2$ is used.  
This parameterization is the sum of two independent Maxwell-Boltzmann thermal 
functions, each term of which can be expressed in terms of the measured coordinates, 
({\mtm},y), integrated over azimuth as 
\begin{equation}
{ {1 \over 2\pi m_{t}} {d^{2}N \over dm_{t}dy} = A(y) m_{t}
e^{-(m_{t}-m_{0})/T_{eff}(y)}}
\label{eq:Thermal with A}
\end{equation}
where the amplitude A and inverse slope parameter, T$_{eff}$ are parameters which can
be extracted from fits to the data at each rapidity slice.  Integration of either of 
these two distributions over m$_t$ produces its contribution to the total number of 
particles per unit of rapidity in the given rapidity slice\cite{Schn93}.
\begin{equation}
{dN \over dy}(y) = 2 \pi A(y) (m_{0}^{2} T_{eff}(y) + 2 m_{0}
T_{eff}^{2}(y) + 2 T_{eff}^{3}(y))
\label{eq:dndy from mtfit}
\end{equation}
Note that A(y) in Eq.\ (\ref{eq:Thermal with A}) can be re-written in terms of 
dN/dy(y), using Eq.\ (\ref{eq:dndy from mtfit}).   

The low-p$_t$ enhancement observed in the pion yields has been attributed to 
feed-down from late-stage resonance decays \cite{Soll90,Hofm95,Wein98}, which tend 
to populate lower pion p$_t$ in the frame of the collision due to the nature of the 
decay kinematics.  In particular, at the beam energies presented here, the delta 
resonances, (such as the 
$\Delta(1232)$) are the predominant mechanism for pion production, due to the very 
large cross-section for pion-nucleon interactions.  The observed asymmetry in the 
spectral shapes of positive and negative pions at low {\mtm} has been described elsewhere by a 
final-state Coulomb interaction of the pions with the nuclear fireball 
\cite{Gyul81,Barz98,Ayal99,Cebr02b}.

The two-slope model which has been applied to the pion spectra was chosen to provide 
the simplest phenomenological description of the data.  Since this parameterization 
does reasonably well at describing the observed shapes of the spectra over all {\mtm}, 
it is used to extract the 4$\pi$ yields of pions in these collisions with minimal 
extrapolation.
\begin{equation}
{1 \over 2\pi m_{t}} {d^{2}N \over dm_{t}dy} = { { dN/dy_{1} m_{t} e^{-(m_{t}-m_{0})/T_{1}} } \over {2\pi
T_{1}  \left( m_{0}^{2} + 2 m_{0} T_{1} + 2 T_{1}^{2} \right) } } + 
{ { dN/dy_{2} m_{t} e^{-(m_{t}-m_{0})/T_{2}} } \over {2\pi
T_{2}  \left( m_{0}^{2} + 2 m_{0} T_{2} + 2 T_{2}^{2} \right) } }
\label{eq:full mtfit}
\end{equation}
The four parameter fit includes two independently fit inverse slope parameters 
(T$_{1}$(y), T$_{2}$(y)), and two independently fit yield parameters (dN/dy$_1$(y), 
dN/dy$_2$(y)), which reasonably describe the low-({\mtm}) and high-({\mtm}) portions 
of the spectrum at each rapidity.  These fits are shown as solid lines in Figs. 
\ref{fig:pim_mt} and \ref{fig:pip_mt}.  The data at more forward rapidities, where in 
particular the {\pip} spectra are limited to the range {\mtm} $<$ 200 MeV/c, are not included in 
the two-slope parameterization fits.

\begin{figure}
\resizebox*{\FigFactor\textwidth}{!}{\includegraphics{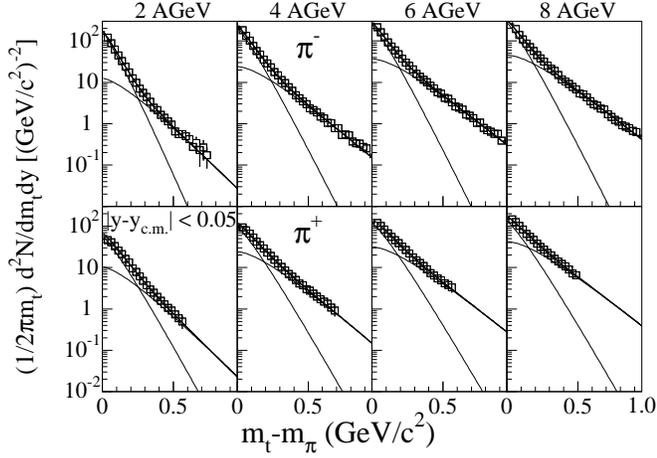}}
\caption{Pion transverse mass spectra at mid-rapidity for 2, 4, 6, and 8 AGeV 
(uppermost curves on Figs. \ref{fig:pim_mt} and \ref{fig:pip_mt}).  Two-slope thermal 
fits are shown superposed on the data along with the contributions to the total from 
the two separate inverse slope terms in Eq. (\ref{eq:full mtfit}).}
\label{fig:pion_midy}
\end{figure}

\begin{figure}
\resizebox*{\FigFactor\textwidth}{!}{\includegraphics{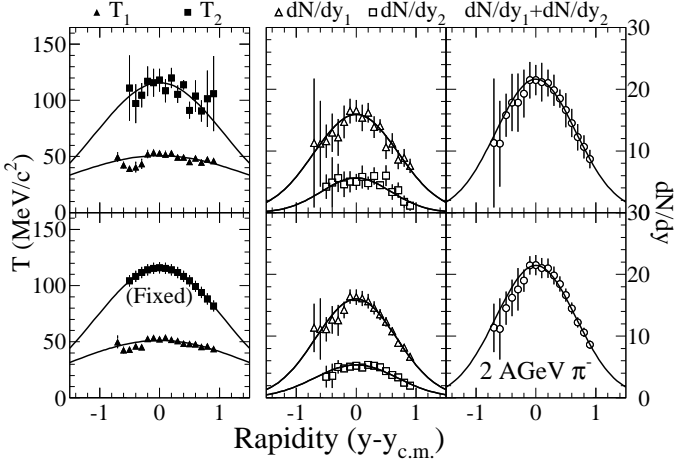}}
\caption{Rapidity dependence of the fit parameters from 2 AGeV {\pim} {\mtm} spectra 
fits with a two-slope model.  The leftmost panels show the inverse slope parameters.  
T$_1$ (indicated by closed triangles) dominates the low-({\mtm}) range of the spectra, 
whereas T$_2$, (indicated by closed squares) dominates the high-({\mtm}) end of the 
spectra.  The middle panels show the two dN/dy fit parameters, (dN/dy$_1$ in open 
triangles and dN/dy$_2$ in open squares) and the rightmost panels show the total dN/dy, 
which is the sum of dN/dy$_1$ and dN/dy$_2$.  The top row reports the fit results when 
all four parameters are allowed to be independent in the fit.  The bottom row shows 
the results for the remaining three parameters when the high-({\mtm}) inverse slope is 
fixed to the Gaussian estimation.  Note that the total dN/dy is fairly 
insensitive to the interplay among the parameters. }
\label{fig:2gev_pim_params}
\end{figure}

\begin{figure}
\resizebox*{\FigFactor\textwidth}{!}{\includegraphics{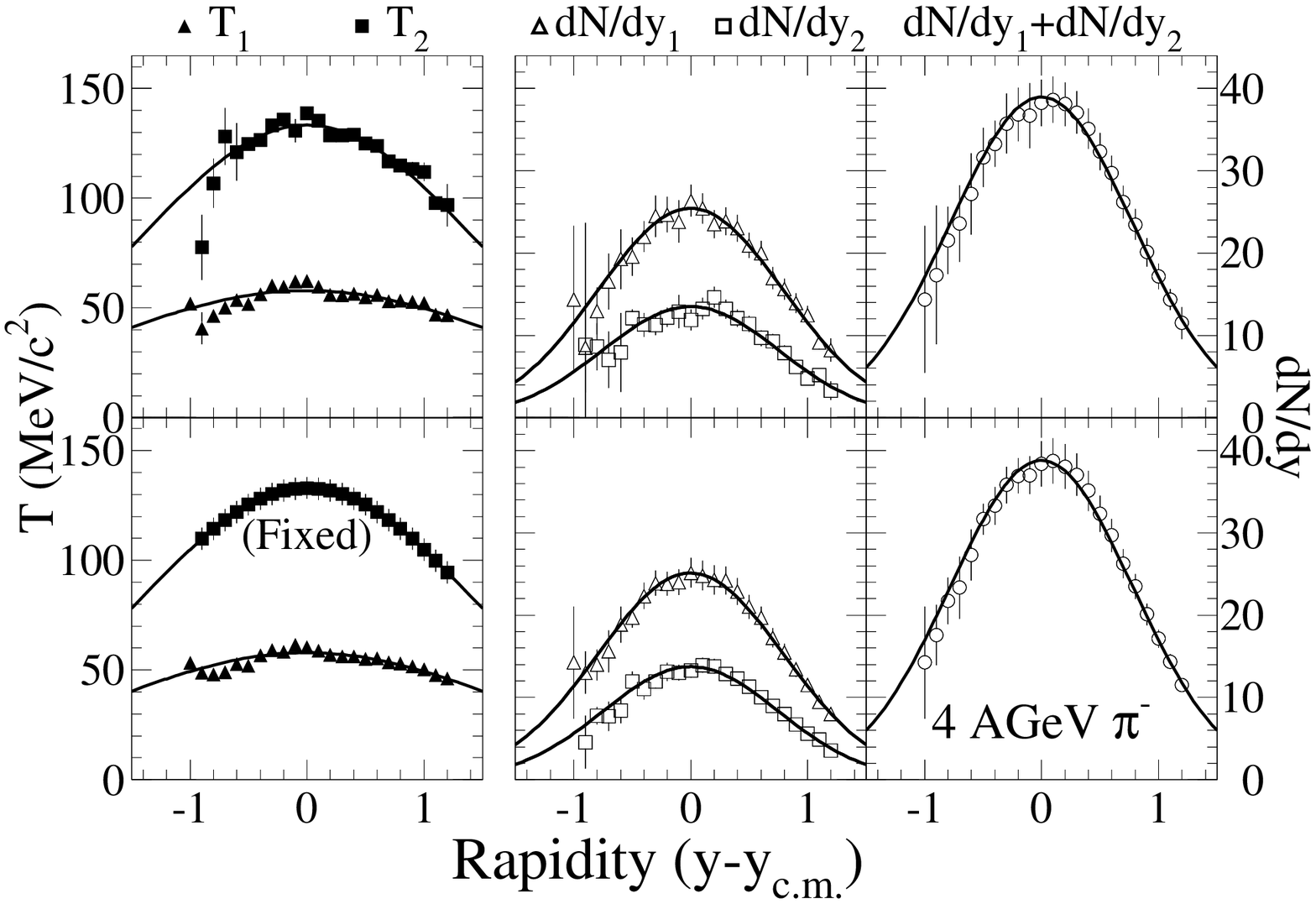}}
\caption{Rapidity dependence of the fit parameters from 4 AGeV {\pim} {\mtm} spectra 
fits with a two-slope model.  See caption of Fig.~\ref{fig:2gev_pim_params} for details.}
\label{fig:4gev_pim_params}
\end{figure}

\begin{figure}
\resizebox*{\FigFactor\textwidth}{!}{\includegraphics{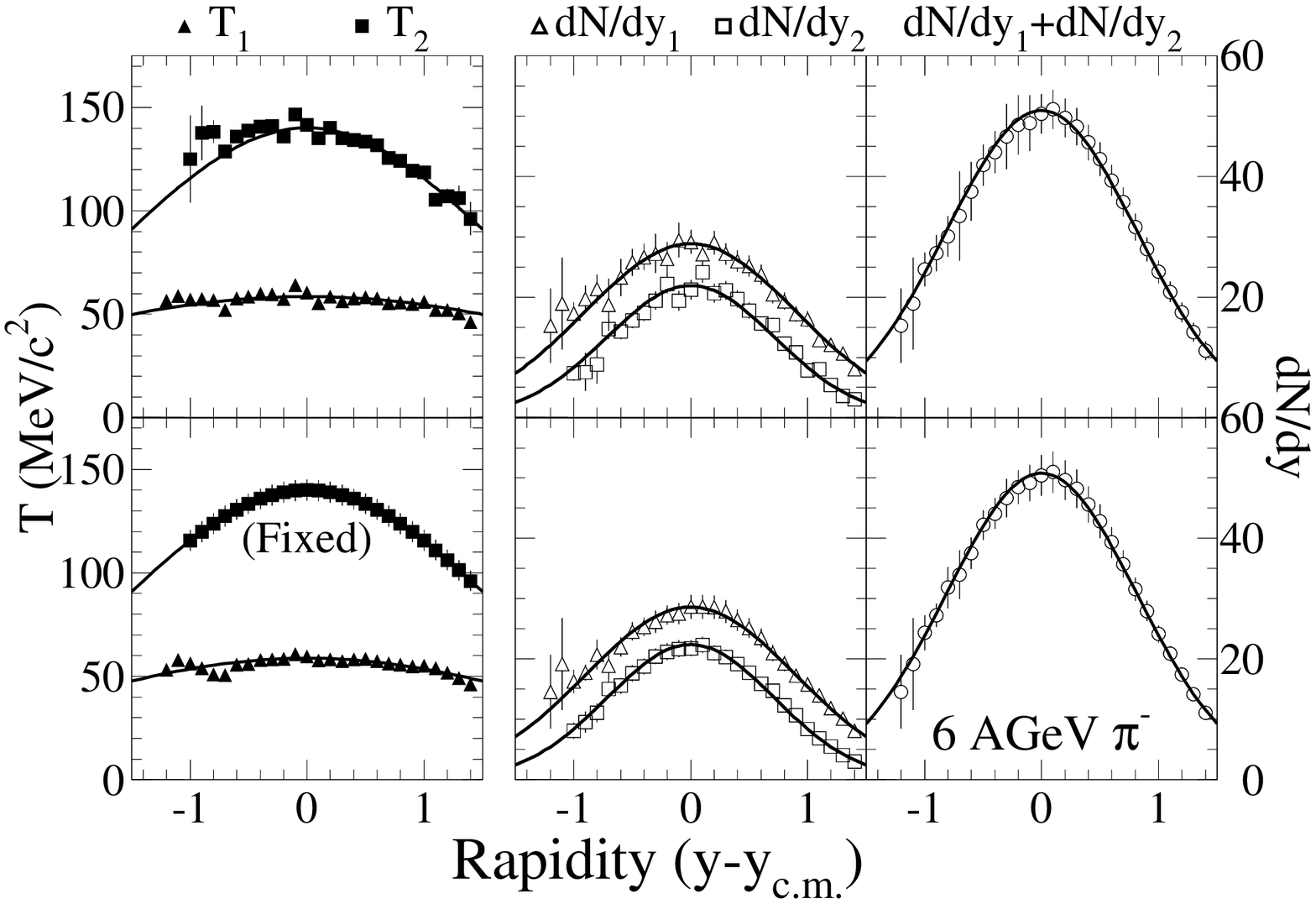}}
\caption{Rapidity dependence of the fit parameters from 6 AGeV {\pim} {\mtm} spectra 
fits with a two-slope model.  See caption of Fig.~\ref{fig:2gev_pim_params} for details.} 
\label{fig:6gev_pim_params}
\end{figure}

\begin{figure}
\resizebox*{\FigFactor\textwidth}{!}{\includegraphics{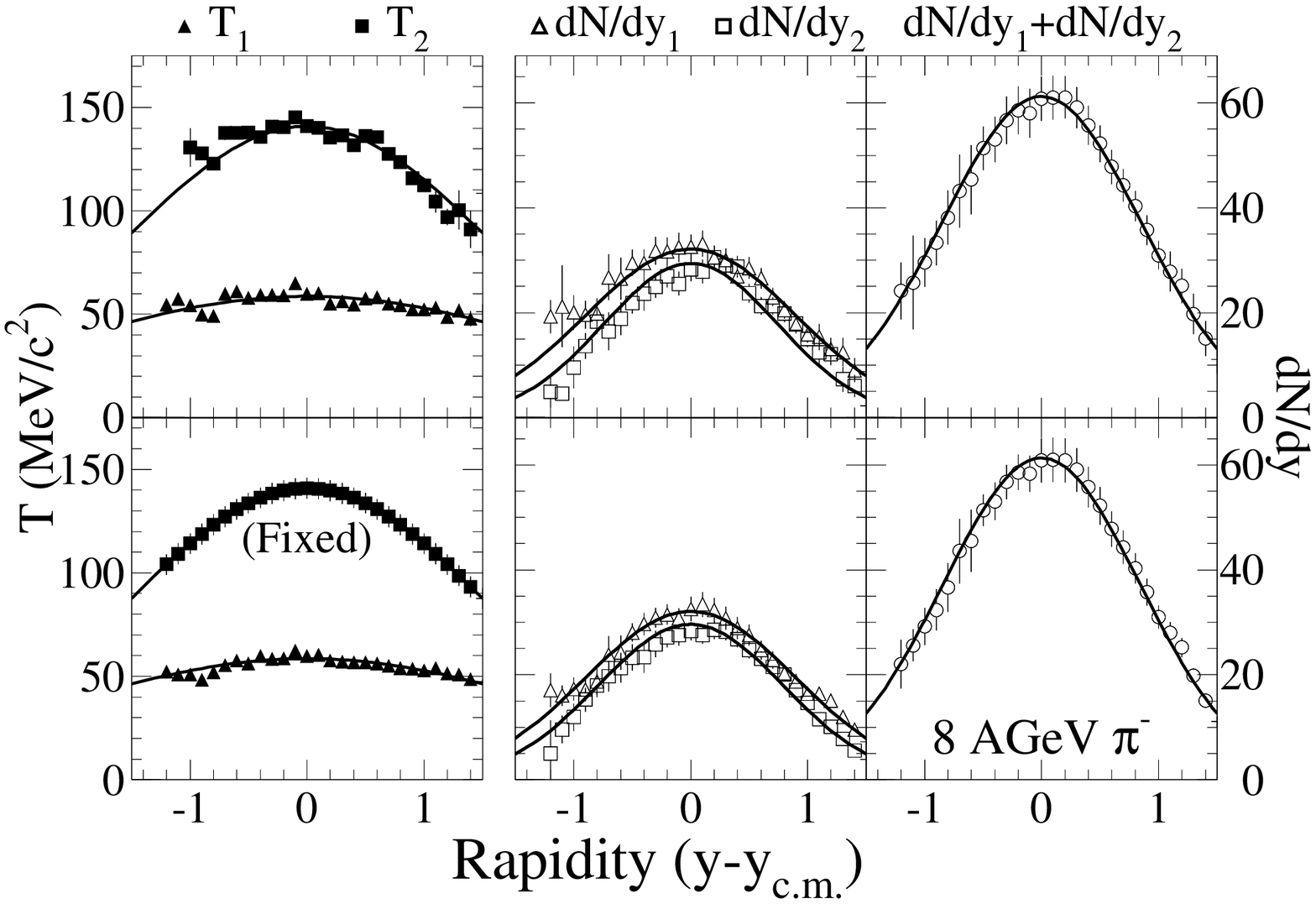}}
\caption{Rapidity dependence of the fit parameters from 8 AGeV {\pim} {\mtm} spectra 
fits with a two-slope model.  See caption of Fig.~\ref{fig:2gev_pim_params} for details.} 
\label{fig:8gev_pim_params}
\end{figure}

More detailed examples of these two-slope fits at mid-rapidity are shown in 
Fig.~\ref{fig:pion_midy}. 
Figs.~\ref{fig:2gev_pim_params}-\ref{fig:8gev_pim_params} show the four fit 
parameters as a function of rapidity from the {\pim} spectra and the resulting 
total dN/dy.  The parameters plotted in the top row are from fits in which 
all four parameters are allowed to be free.  All of them show an approximately 
Gaussian dependence on rapidity, which is demonstrated by the solid line in each panel.
The bottom row of each figure demonstrates the result of a second fit of the spectra 
in which the high-({\mtm}) inverse slope parameter is constrained to the Gaussian 
(solid line) value.  This procedure smooths out the covariance between the individual 
yield parameters but has a negligible effect on total dN/dy (rightmost panel).  

\begin{figure}
\resizebox*{\FigFactor\textwidth}{!}{\includegraphics{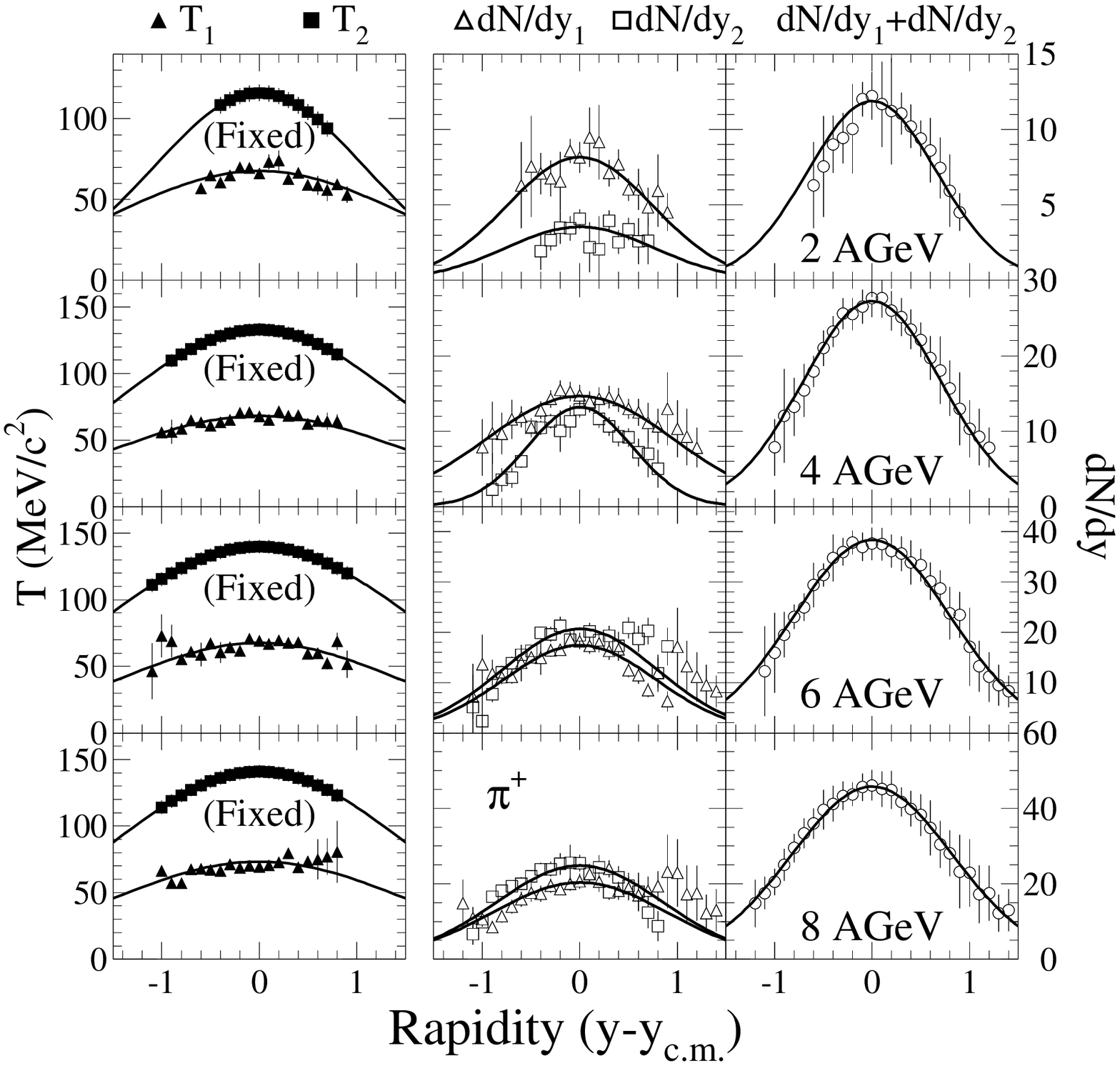}}
\caption{Rapidity dependence of the fit parameters for each beam energy from the {\pip} {\mtm} spectra 
fits with a two-slope model.  (See caption of Fig.~\ref{fig:2gev_pim_params} for more details.)   The 
high-({\mtm}) inverse slope parameters have been fixed to the {\pim} parameterized values.}
\label{fig:all_pip_params}
\end{figure}

If the source of pions were a static thermal source of temperature T$_0$ and zero 
chemical potential, the expected shape of T$_{eff}$ in Figs.  
\ref{fig:2gev_pim_params}-\ref{fig:all_pip_params} would be T$_{0}/\cosh(y)$. (Of 
course, there would then be needed only a single term in Eq. (\ref{eq:full mtfit}).)  
We have shown in Ref. \cite{Klay02} that there is significant longitudinal flow - {\it 
i.e.}, the source is not static.  As the main purpose of the fits in Figs. 
\ref{fig:pim_mt} and \ref{fig:pip_mt} is for integration of the ({\mtm},y) spectra to 
obtain 4$\pi$ yields, a Gaussian fit of the inverse slope parameters in Figs.  
\ref{fig:2gev_pim_params}-\ref{fig:all_pip_params}, rather than 1/$\cosh(y)$, was used.

Fig.~\ref{fig:all_pip_params} shows the {\pip} fit parameters at each beam energy.  
The limited range in {\mtm} of the $\pi^+$ spectra is caused by the {\meandedx} 
overlap with the protons, which makes it impossible to get good spectral data at all 
rapidity slices.  In order to make reasonable estimates of the 4$\pi$ positive pion 
yields, the high-({\mtm}) positive pion inverse slope parameters were assumed to be 
the same as those of the negative pions\cite{Pelt97}.  In the estimation of the 
$\pi^+$ dN/dy, the high-({\mtm}) inverse slopes were therefore fixed to the 
negative pion values.  Experimentally, the {\mtm} negative pion inverse slope 
parameters do reasonably reproduce the observed high-({\mtm}) positive pion inverse 
slopes in the rapidity regions where they can be measured.  Where the {\mtm} 
spectra are severely truncated, the positive pion yields were obtained from 
single-slope fits.  The reported systematic uncertainties in these rapidity slices are 
correspondingly larger to account for this missed yield.

\begin{figure}
\resizebox*{\FigFactor\textwidth}{!}{\includegraphics{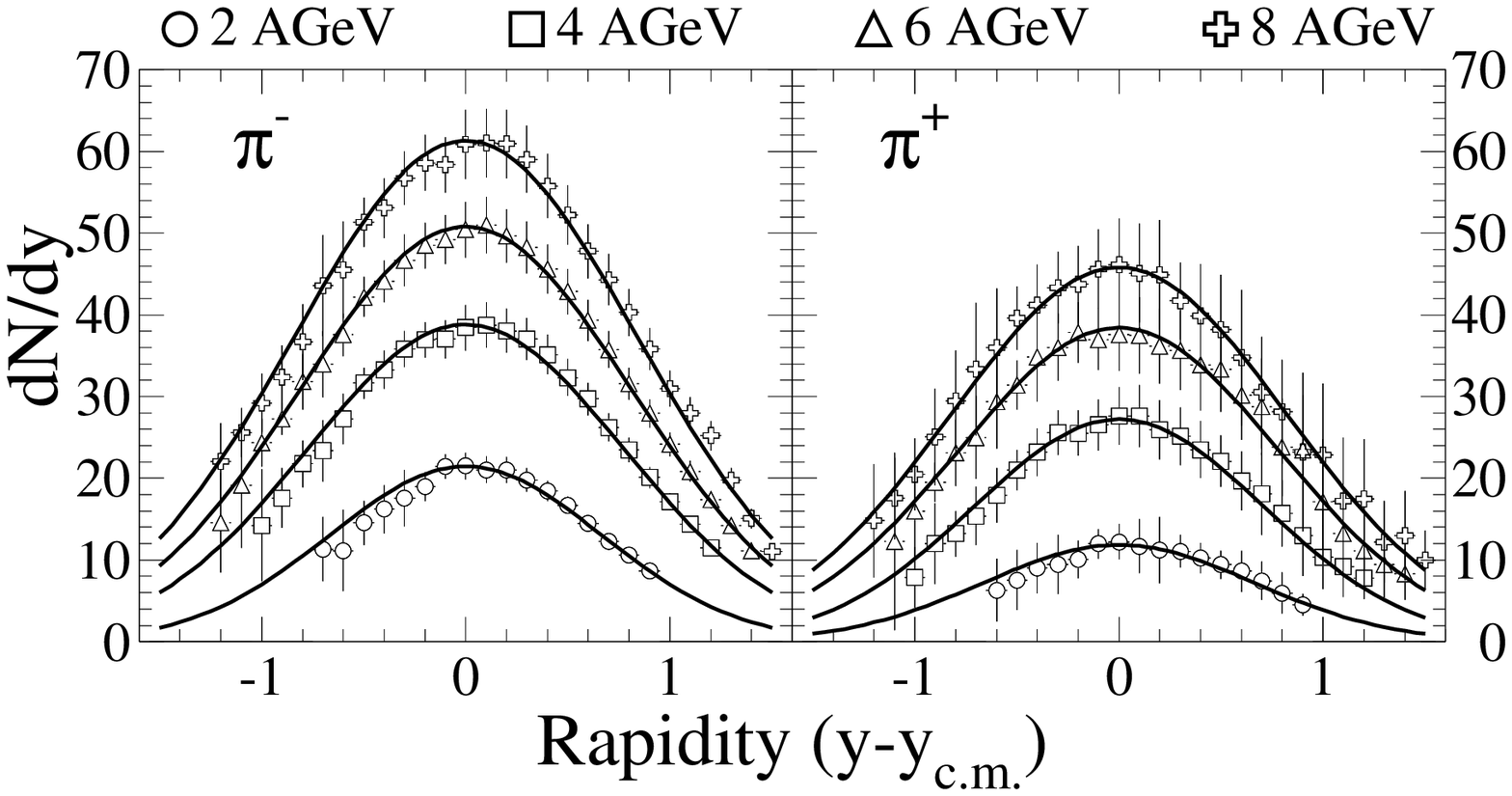}}
\caption{Yield per event of {\pim} (left-hand panel) and {\pip} (right-hand panel) in 
0-5\% central Au+Au collisions at 2, 4, 6, and 8 AGeV integrated over {\mtm} as a 
function of rapidity.  A Gaussian parameterization is applied to extract 
the 4$\pi$ yields.  Reported uncertainties are statistical plus systematic.}
\label{fig:pion_dndy}
\end{figure}

Fig.~\ref{fig:pion_dndy} shows the beam energy dependence of the rapidity 
distributions of negative and positive pions extracted from the fits.  These distributions are 
well-described by Gaussians, which are used to obtain the total 4$\pi$ yields by integrating the 
fitted distributions over all rapidity.
\begin{equation}
\langle \pi \rangle = \int^{+\infty}_{-\infty} {dN \over dy} dy = \sqrt{2\pi}wQ
\end{equation}
where Q is the value of dN/dy at y-y$_{CM}$ = 0 and w is the width.
The fit parameters and their uncertainties are listed in Tables \ref{PIM
dNdy Parameters} and \ref{PIP dNdy Parameters}.  Both the widths and the overall 
yields of pions increase as a function of beam energy.  However, there is a more 
significant increase in the observed pion yields between 2 and 4 AGeV than between 4 
and 6 AGeV or 6 and 8 AGeV.  

\begin{table}
{\centering \begin{tabular}{cccc}
E$_{beam}$ & Q$_{\pi^-}$ & $\langle \pi^{-} \rangle$ & w$_{\pi^{-}}$\\ \hline 
2 AGeV & 21.3 $\pm$ 0.1 $\pm$ 1.3 & 36.1 $\pm$ 0.3 $\pm$ 2.0 & 0.675 $\pm$ 0.006 $\pm$ 0.005 \\
4 AGeV & 39.0 $\pm$ 0.1 $\pm$ 2.1 & 76.0 $\pm$ 0.2 $\pm$ 4.2 & 0.780 $\pm$ 0.003 $\pm$ 0.002 \\
6 AGeV & 50.8 $\pm$ 0.1 $\pm$ 2.7 & 104.0 $\pm$ 0.2 $\pm$ 5.4 & 0.817 $\pm$ 0.002 $\pm$ 0.001 \\
8 AGeV & 61.1 $\pm$ 0.1 $\pm$ 3.3 & 130.7 $\pm$ 0.4 $\pm$ 7.9 & 0.854 $\pm$ 0.003 $\pm$ 0.008 \\
\end{tabular}\par}  
\caption{dN/dy fit parameters for negative pions at each beam energy with statistical and systematic uncertainties 
reported separately.  Q$_{\pi^-}$ is the amplitude of the distribution at y-y$_{CM}$ = 0, 
$\langle \pi^{-} \rangle$ is the 4-$\pi$ yield (total area) and 
w$_{\pi^{-}}$ is the width.}
\label{PIM dNdy Parameters} \end{table}

\begin{table}
{\centering \begin{tabular}{cccc}
E$_{beam}$ & Q$_{\pi^+}$ & $\langle \pi^{+} \rangle$ & w$_{\pi^{+}}$\\
\hline
2 AGeV & 11.5 $\pm$ 0.3 $\pm$ 1.2 & 19.2 $\pm$ 1.3 $^{+2.0}_{-1.0}$ & 0.668 $\pm$ 0.053 $\pm$ 0.011 \\
4 AGeV & 27.7 $\pm$ 0.3 $^{+1.2}_{-1.4}$ & 46.3 $\pm$ 0.8 $^{+4.5}_{-3.3}$ & 0.667 $\pm$ 0.015 $\pm$ 0.025 \\
6 AGeV & 38.4 $\pm$ 0.3 $^{+2.4}_{-1.8}$  & 75.7 $\pm$ 1.1 $^{+3.4}_{-2.9}$ & 0.787 $\pm$ 0.014 $\pm$ 0.016 \\
8 AGeV & 46.2 $\pm$ 0.4 $^{+2.7}_{-3.4}$ & 95.9 $\pm$ 1.1 $^{+6.1}_{-5.9}$ & 0.828 $\pm$ 0.013 $\pm$ 0.005 \\
\end{tabular}\par}  
\caption{dN/dy fit parameters for positive pions at each beam energy with statistical and systematic uncertainties 
reported separately.  Q$_{\pi^+}$ is the amplitude of the distribution at y-y$_{CM}$ = 0,
$\langle \pi^{+} \rangle$ is the 4-$\pi$ yield (total area) and w$_{\pi^{+}}$ is the width.}
\label{PIP dNdy Parameters}
\end{table}

\section{Discussion}

In this section, some of the characteristics of the observed pion spectra obtained 
from E895 are discussed.  The pion spectral shapes are compared to the predictions of 
a microscopic transport model, the dN/dy distributions are evaluated in the context of 
collective dynamics, and the overall yields are used to infer the initial state 
entropy density obtained in these collisions.

It is interesting to note the asymmetry between the positive and negative pion yields 
at each beam energy.  Although the ratio of negative to positive pions decreases over 
the studied beam energy range, from 1.88 at 2 AGeV to 1.36 at 8 AGeV (see Tables 
\ref{PIM dNdy Parameters} and \ref{PIP dNdy Parameters}), it does not reach the 
asymptotic value of $\sim$ 1.0 observed at the top CERN SPS energy\cite{Dunn97,NA4902}.  
The neutron excess in Au+Au collisions (118+118 neutrons compared to 79+79 protons) 
combined with the pion branching ratios suggests that 1.95 negative pions for every 
positive pion will be produced \cite{Stoc86}.  If the isospin fractions are folded 
with the observed cross-sections for NN $\rightarrow$ NN$\pi$ from experimental 
measurements\cite{VerW82}, the expected ratio of 
$\langle\pi^{-}\rangle$:$\langle\pi^{+}\rangle$ at 1 AGeV is $\sim$ 1.91:1.  The 2 
AGeV data, with $\sqrt{s_{NN}}$ not far from the $\Delta$(1232) production threshold, are 
quite near, though a little lower than this predicted ratio.  As the energy increases, 
the number of directly produced pion pairs ($\pi^-\pi^+$) is expected to lower the 
ratio, asymptotically approaching 1.0, which is the trend we observe.  At 8 AGeV the 
negative pion excess is only approximately 36\%, compared to 88\% at 2 AGeV.

\subsection{RQMD Comparisons}

RQMD (Relativistic Quantum Molecular Dynamics) v2.3 \cite{Sorg95} is a microscopic 
transport model which attempts to simulate heavy ion collisions by propagating 
individually all particles through the six dimensions of phase space in the fireball.  
Interaction probabilities are approximated by using published interaction 
cross-sections of free hadrons and the relative phase space proximity of pairs of 
particles at each time step of the reaction.   Inelastic collisions may produce new 
particles, such as pions, which are also propagated through phase space along with the 
nucleons.  The reaction ends when the phase space density reaches a low enough 
threshold such that the probability of further interactions is small - the freeze-out 
point.  This model has been reasonably successful in describing many final state 
observables experimentally measured in the beam energy range studied for this analysis 
\cite{Bass98}.

In this and other cascade models, final-state particle distributions are frozen at the 
end of the reaction.  Post freeze-out effects, such as the Coulomb interaction of the 
pions with the nuclear fireball, which will be discussed in detail using results from 
this analysis elsewhere \cite{Cebr02b}, are not included.  However, RQMD combined with 
an afterburner to simulate final state Coulomb interactions \cite{Xu96} and to  
permit the decay of residual resonances has been successful at describing 
asymmetries in observed pion yields at low p$_{t}$.

\begin{figure}
\resizebox*{\FigFactor\textwidth}{!}{\includegraphics{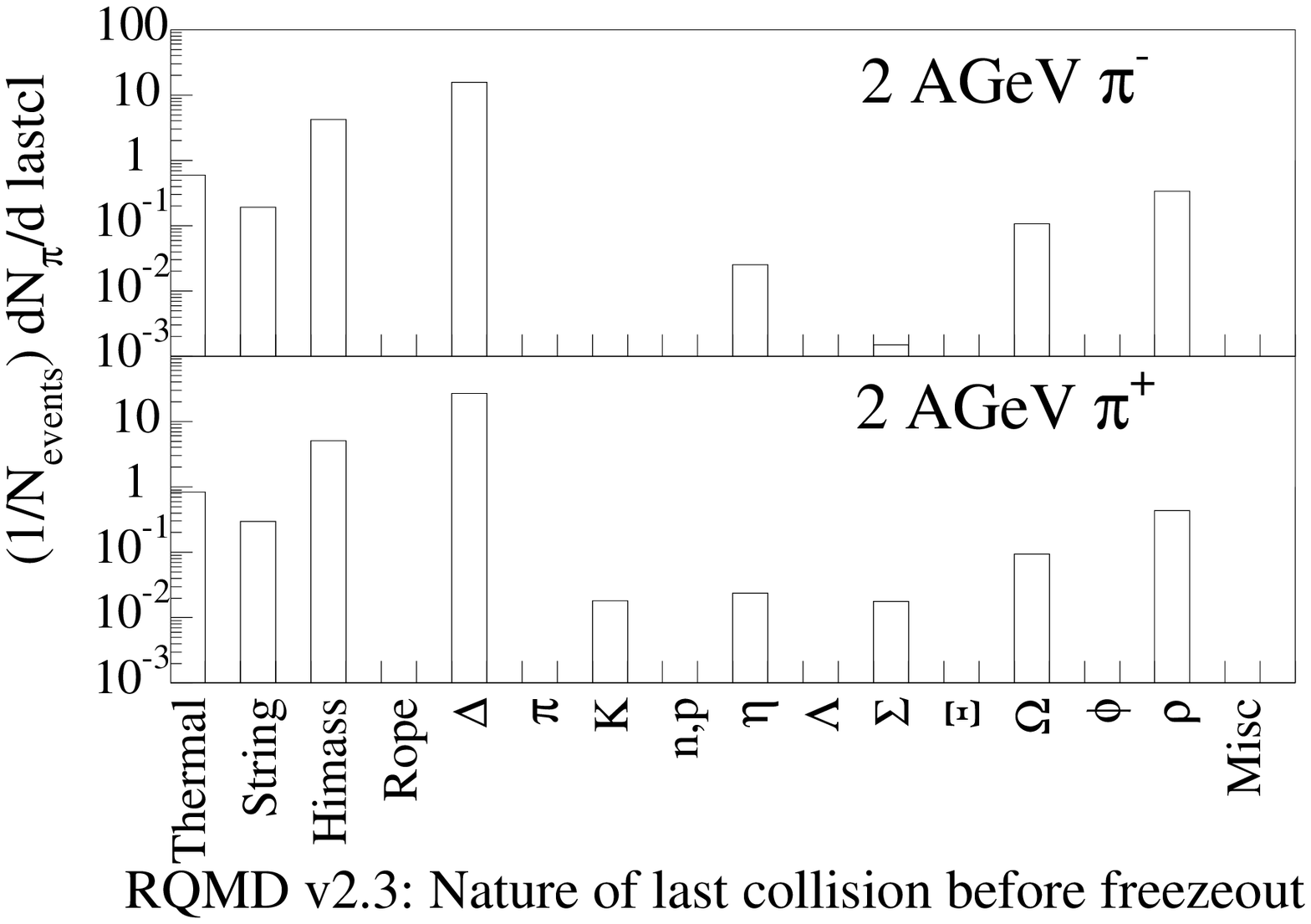}} 
\caption{RQMD nature of pion last collision, ``lastcl'', before freeze-out at 2 AGeV 
for impact parameter, b $\leq$ 3 fm.} 
\label{fig:rqmd_lastcl_2gev} 
\end{figure}

\begin{figure}
\resizebox*{\FigFactor\textwidth}{!}{\includegraphics{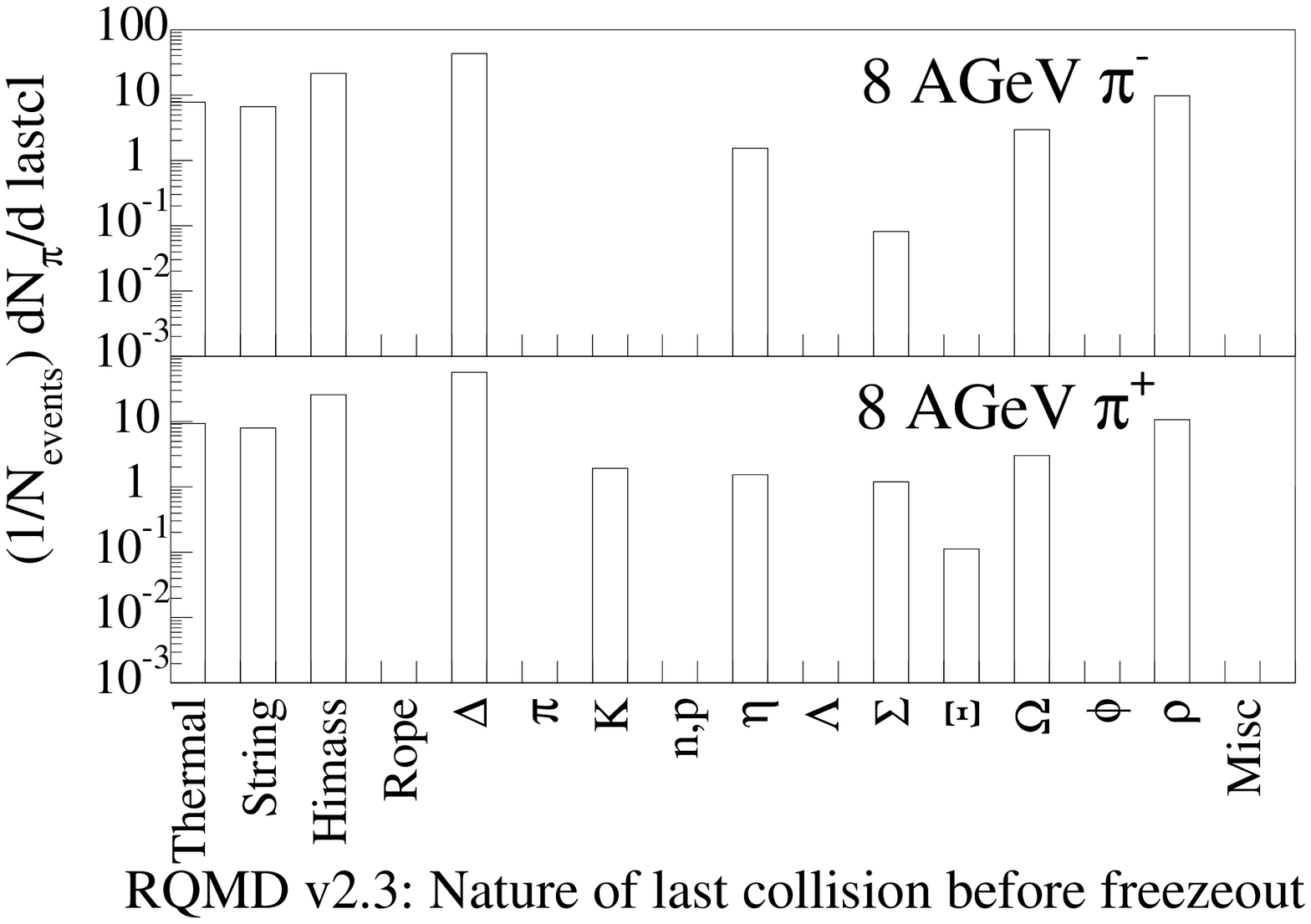}} 
\caption{RQMD nature of pion last collision, ``lastcl'', before freeze-out at 8 AGeV 
for impact parameter, b $\leq$ 3 fm.} 
\label{fig:rqmd_lastcl_8gev} 
\end{figure}

RQMD version 2.3, with the nucleon mean field setting turned on, was used to generate 
pion distributions from central (b $\leq$ 3 fm) Au+Au collisions for comparison with 
the data obtained by E895 for this paper.  One of the output parameters from RQMD 
records the nature of the last collision, ``lastcl'', of each particle before 
freeze-out.  Particles whose lastcl was a thermal rescattering are labelled 
``thermal''.  Particles whose lastcl was the decay of a resonance have the parent 
particle listed explicitly.  However, high mass resonances which are not Deltas or 
vector mesons are combined in a single category labelled ``himass'' in the user's 
notes for the code.  Particles whose lastcl was String or Rope fragmentation are 
indicated separately.  Figs. \ref{fig:rqmd_lastcl_2gev} and \ref{fig:rqmd_lastcl_8gev} 
show the nature of the last collision for {\pim} and {\pip} for a centrality 
selection of b $\leq$ 3 fm at 2 AGeV and 8 AGeV, respectively.  The distributions 
are normalized to the per event yield of pions.  In each case, the Delta resonances 
are the largest single contributor to the last pion interactions before freeze-out.  

\begin{figure}
\resizebox*{\FigFactor\textwidth}{!}{\includegraphics{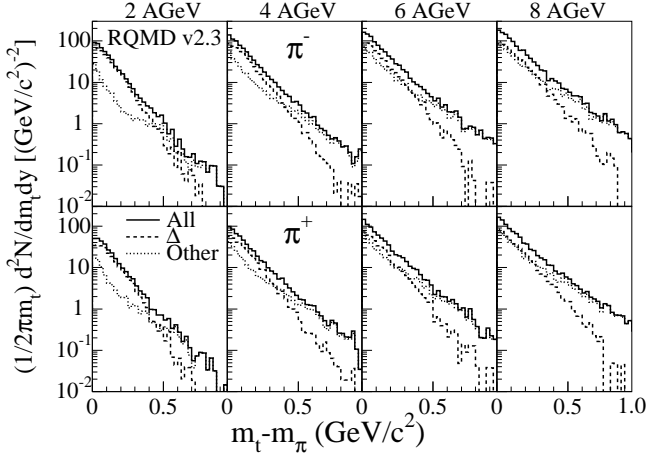}}
\caption{RQMD charged pion spectra at mid-rapidity for collisions with impact
parameter, b $\leq$ 3 fm.  The contribution to the spectra from pions whose last
interaction was a Delta decay and the sum of all other processes are shown 
separately.} 
\label{fig:rqmd_midy}
\end{figure}

\begin{figure}
\resizebox*{\FigFactor\textwidth}{!}{\includegraphics{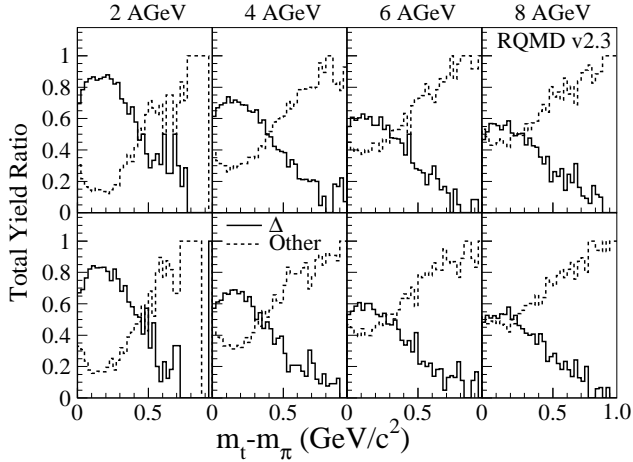}}
\caption{Ratios of pion yield for different production mechanisms to the total 
yield as a function of {\mtm} at mid-rapidity from RQMD collisions with impact 
parameter, b $\leq$ 3 fm.  The contribution to the spectra from pions whose last 
interaction was a Delta decay and from the sum of all other processes are shown 
separately.} 
\label{fig:rqmd_ratios}
\end{figure}

The dependence of the pion spectral shapes on feed-down from Delta resonances is 
evident in Fig.~\ref{fig:rqmd_midy}, which shows the {\mtm} spectra of pions near
mid-rapidity ($|$y$|$ $<$ 0.3) from RQMD.  Pions whose last collision was a Delta
are plotted separately from the pions with all other last interactions.  The
spectral shapes are strongly affected by a Delta lastcl.  Plotted in  
Fig.~\ref{fig:rqmd_ratios} are the ratios of the pion yields from Deltas and all 
other last interactions to the total yields as a function of {\mtm}.  In all cases 
the Delta pions contribute more significantly to the total yield at low {\mtm} 
than at high {\mtm}, but the influence of the Delta contribution to the spectra 
diminishes as the beam energy increases.  The total mid-rapidity pion spectra are 
fit with the same model as was used to describe the data, Eq. (\ref{eq:full 
mtfit}), with the fit parameters shown in Tables \ref{table:PIM RQMD FITS} and 
\ref{table:PIP RQMD FITS}.  At both high and low {\mtm}, the RQMD {\pim} and 
{\pip} appear to have common inverse slope parameters and there is no obvious 
evolution with beam energy.  The average value at low {\mtm} (T$_1$) is $\sim$72 
MeV, while the high {\mtm} (T$_2$) average is $\sim$140 MeV.  The magnitude of 
the low {\mtm} RQMD inverse slope parameters is much larger than what is observed 
in the data at all beam energies.  At high {\mtm}, the RQMD inverse slope 
parameters are much closer to the observed values, except at 2 AGeV, where the 
data show T$_2 \sim$ 120 MeV at mid-rapidity. 

\begin{table}
{\centering \begin{tabular}{ccccc}
E$_{beam}$ (AGeV) & dN/dy$_{1}$ & T$_{1}$ (MeV/c$^2$) & dN/dy$_{2}$ & T$_{2}$ (MeV/c$^2$)\\
\hline
2 & 17.4 & 73 & 1.8 & 146\\
4 & 21.3 & 73 & 10.9 & 131\\
6 & 26.6 & 74 & 15.5 & 146\\
8 & 31.7 & 77 & 20.1 & 155\\
\end{tabular} \par}
\caption{Mid-rapidity RQMD {\pim} fit parameters, Eq. (\ref{eq:full mtfit}).}
\label{table:PIM RQMD FITS}
\end{table}

\begin{table}
{\centering \begin{tabular}{ccccc}
E$_{beam}$ (AGeV) & dN/dy$_{1}$ & T$_{1}$ (MeV/c$^2$) & dN/dy$_{2}$ & T$_{2}$ (MeV/c$^2$)\\
\hline
2 & 10.2 & 73 & 1.6 & 134\\
4 & 14.4 & 72 & 8.5 & 131\\
6 & 15.5 & 64 & 17.7 & 131\\
8 & 21.4 & 70 & 21.3 & 144\\
\end{tabular} \par}
\caption{Mid-rapidity RQMD {\pip} fit parameters, Eq. (\ref{eq:full mtfit}).}  
\label{table:PIP RQMD FITS}
\end{table}

\begin{figure}
\resizebox*{\FigFactor\textwidth}{!}{\includegraphics{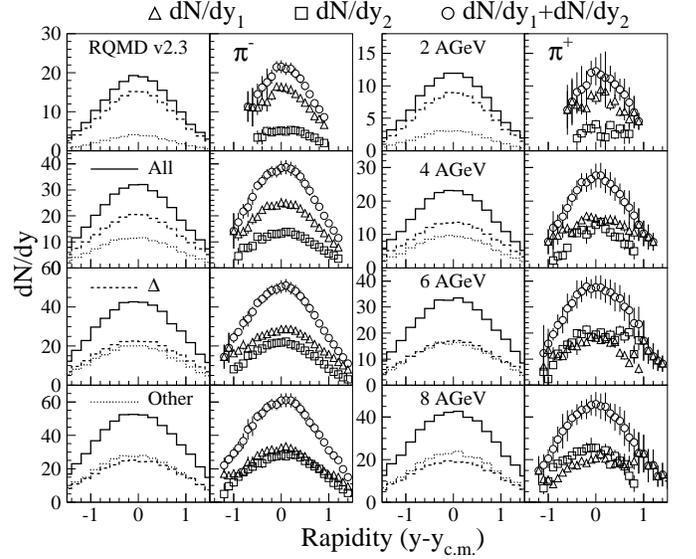}}
\caption{Rapidity density of charged pions from RQMD calculations for collisions with 
impact parameter, b $\leq$ 3 fm.  The contribution from Delta resonance 
decay pions and the sum of all other processes are also plotted separately
to show the relative contribution to the pion yields from Delta decays.  Also 
plotted are the data from Figs.~\ref{fig:2gev_pim_params}-\ref{fig:all_pip_params}.}
\label{fig:rqmd_dndy}
\end{figure}

The rapidity density distributions of charged pions predicted by RQMD are shown in 
Fig.~\ref{fig:rqmd_dndy} alongside the E895 data from 
Figs.~\ref{fig:2gev_pim_params}-\ref{fig:all_pip_params}.  
The contributions to the total dN/dy from Delta resonance decay pions and all 
other pions are also shown.  As the beam energy increases, the contribution to 
the total from Delta resonance decay pions becomes less important, which is 
consistent with the trends seen in the data.

\subsection{Longitudinal flow}

An important question in heavy ion collisions is the degree of collectivity of the 
produced particles, which can arise from the build-up of pressure in the hot, dense 
collision zone.  RQMD has no such hydrodynamic effect explicitly included.  However, individual 
particle thermal rescattering, the microscopic analog to pressure, can also drive collective flow, 
and has been observed in RQMD calculations at RHIC energies \cite{Monr99}.  
Rapidity broadening along the beam axis has been observed in heavy ion 
collisions at many beam energies for a wide range of systems and particle species 
\cite{Schn93,Stac96,Brau95,Klay02}.  This broadening has been interpreted as arising from the 
collective motion of the system after collision in the longitudinal direction (longitudinal flow). 
The observed rapidity distributions are compared with the expectation for a stationary thermal 
source, and with a longitudinally boost-invariant superposition of multiple boosted individual
isotropic, locally thermalized sources in a given rapidity interval.  Each locally 
thermalized source is modelled by the {\mt}-integrated Maxwell-Boltzmann distribution, 
Eq. (\ref{eq:Thermal with A}), with the rapidity dependence of the energy, $E = 
m_t\cosh(y)$ explicitly included, and T is true temperature: 
\begin{equation}
{dN_{th} \over dy}(y) = B T^{3} ({{m^2 \over T^2} + {m \over T} {2
\over \cosh y} + {2 \over \cosh^2 y}})
e^{({- {m \over T} \cosh y})}.
\label{eq:Thermal dNdy}
\end{equation}

The distributions are integrated over source element rapidity to extract the maximum 
longitudinal flow, $\eta_{max}$
\begin{eqnarray}
\label{eq:longflow}
{dN \over dy} &=& {\intop}^{\eta_{max}}_{\eta_{min}} d\eta {dN_{th}
\over dy}{(y - \eta)} \\
{\beta_{L}} &=& {\tanh (\eta_{max}) }. \nonumber
\end{eqnarray}
where $\eta_{max}$ = - $\eta_{min}$, from symmetry about the center of mass, and 
$\beta_{L}$ is the maximum longitudinal velocity in units of c.  An average 
longitudinal flow velocity can be defined as $\langle\beta_{L}\rangle = tanh(\eta_{max}/2)$.

Ref. \cite{Klay02} presents longitudinal flow velocities extracted from the proton 
rapidity densities measured by E895 for the same event selection as the pions presented 
in this analysis and compares the results to values extracted from a wide array of 
experiments over the beam energy range from 1 to 160 AGeV.  Since the protons are 
present before the collision, at higher beam energies they may be strongly affected by 
nuclear transparency, which can also broaden the rapidity distributions.  Therefore, 
the protons alone cannot be used to determine the absolute magnitude of the collective 
motion, if it is present.  

In order to determine the degree of collectivity, it is important to compare multiple 
particle species from the same collisions simultaneously.  This has been done with 
central Si+Al collisions at 14.6 AGeV for protons, pions, kaons and lambda hyperons 
\cite{Brau95} and in central S+S collisions at 200 AGeV for pions, kaons and lambda 
hyperons \cite{Schn93}.  A common collective flow velocity was able to reasonably 
describe all of the observed rapidity distributions, except the protons in S+S collisions at 200 
AGeV, where apparent nuclear transparency is more pronounced.  A similar simultaneous description 
is possible here, by combining the proton rapidity distributions from \cite{Klay02} with 
the present pion rapidity distributions.  

\begin{figure}
\resizebox*{\FigFactor\textwidth}{!}{\includegraphics{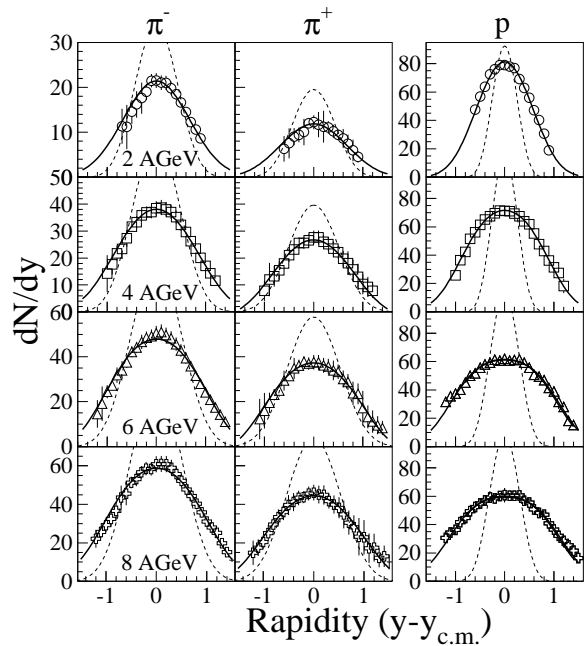}} 
\caption{$\pi^-$, $\pi^+$ and proton rapidity density distributions at
2,4,6, and 8 AGeV.  The expectation for isotropic emission from a
stationary thermal source is indicated with dashed lines, whereas the solid
curves represent the form including longitudinal flow from the proton $\eta_{max}$ 
values given in Ref. \cite{Klay02}.  In all cases the thermal model is too narrow.  The 
proton data and the longitudinal expansion velocity come from Ref. \cite{Klay02}.}
\label{fig:all_longflow} 
\end{figure}

The rapidity densities of pions and the protons from Ref. \cite{Klay02} are shown in 
Figure \ref{fig:all_longflow}.   Stationary thermal source emission functions, the sum 
of two of Eq. (\ref{eq:Thermal dNdy}) for the pions using the two inverse slope 
parameters from the pion transverse mass spectra fits at mid-rapidity, are shown as 
dashed lines.  We have not measured the true system temperature, T, since 
T$_1$ and T$_2$ parametrize the temperature and known radial flow, resonance 
and Coulomb effects in the pion spectra.  Consequently, T$_2$ overestimates 
the system temperature and T$_1$ may underestimate (due to competing effects 
of radial flow, resonance feed-down and the Coulomb interaction with the 
fireball).  Therefore the plotted distributions (dashed lines) are probably wider than the 
true thermal distributions would be, and yet are still too narrow to reproduce the 
observed rapidity spectra.  The inclusion of longitudinal flow remedies
this.  For the protons, a single  Eq. (\ref{eq:Thermal dNdy}) is used for the thermal 
rapidity distribution, as in Ref. \cite{Klay02}.  

The solid curves are the emission functions including longitudinal flow, Eq. 
(\ref{eq:longflow}), with the velocities fixed to the values extracted from the 
protons ($\langle\beta_{L}\rangle$ = 0.28, 0.42, 0.48, 0.50 at 2, 4, 6 and 8 AGeV, 
respectively).  The longitudinally expanding source clearly better reproduces the 
measured pion distributions at all beam energies than do the dashed curves in 
Fig.~\ref{fig:all_longflow}.  Consistency among particles of different masses 
supports a hydrodynamical interpretation of the rapidity density broadening; the 
system is expanding like a fluid with a common longitudinal flow velocity.  

\subsection{Entropy production}
        
        It was suggested many years ago by Fermi \cite{Ferm50} and later 
Landau \cite{Land53} that pion production may be used to estimate the 
amount of entropy produced in high energy particle collisions.  Later,   
Van Hove \cite{VanH82} extended this idea to heavy ion collisions and
proposed that this may be a way to distinguish events in which a Quark
Gluon Plasma (QGP) is formed.  In a QGP the color degrees of
freedom of the liberated partons introduce a significant number of new
energy states unavailable in a hadron gas.  By studying pion production
over a broad range of collision systems and energies, discontinuities in
the observed multiplicities might indicate the onset of QGP formation.

In 1995, Ga\'{z}dzicki \cite{Gazd95a} took the available data from heavy
ion collisions and showed that there is an increase in the observed
entropy produced at the SPS (NA35 Experiment with S+S collisions at 200 AGeV 
\cite{Bach94}) compared with AGS energies.  The low energy heavy ion data follow the 
trend for p+p collisions, while at the SPS there is an apparent factor of 3 increase in 
the effective number of degrees of freedom \cite{Gazd95b}.  The entropy production 
analysis of Pb+Pb collisions at 158 AGeV (NA49) \cite{Brad98} supported this observation.  

This model assumes that the entropy is produced at the early stage of the collision 
when the incident matter is in a highly excited state.  The thermalized, strongly 
interacting matter is assumed to expand adiabatically to the freeze-out point, 
preserving the early stage entropy.  

Since the majority of produced particles are pions, to first order, the mean pion 
(boson) multiplicity should be nearly proportional to the entropy.  The ratio of the 
mean pion multiplicity to the mean number of participating nucleons, $\langle \pi 
\rangle / \langle N_{part} \rangle$, provides a simple estimate of the entropy 
density.  $\langle N_{part} \rangle$ for a nucleus-nucleus collision can be estimated
using a Glauber model calculation of the mean free path of the nucleons through the nuclei as they 
collide at a given impact parameter.  For the present analysis, $\langle$N$_{P}\rangle$ 
was estimated using RQMD.  The nucleons are distributed according to a Woods-Saxon nuclear 
density profile and the impact parameter of the simulated collision is used with the nucleon-nucleon
interaction cross-sections\cite{PDG} to determine the number of participants.  The top 5\% of collisions 
(determined by integrating the $\langle$N$_{P}\rangle$ distribution from a set of minimum bias RQMD 
events at each beam energy) correspond to $\langle$N$_{P}\rangle$ = 364, 366, 365, 363 at 2,4,6, and 8 
AGeV, respectively.  The estimated uncertainty on these values is $\pm$ 5 participants.

\begin{figure}
\resizebox*{\FigFactor\textwidth}{!}{\includegraphics{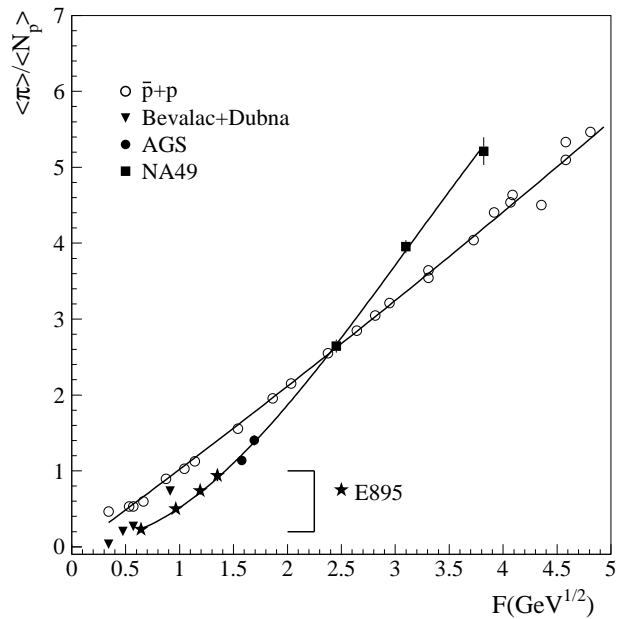}}
\caption{Mean pion multiplicity per participant vs. F, the Fermi energy variable.  The
solid symbols represent data from heavy ion collisions, whereas the
open symbols come from p+p interactions.  The E895 data points are
indicated by stars.  The third-order polynomial fit to the heavy ion data show that a
smooth trend with F is possible, although the low energy and high energy data 
(including RHIC results) can be well-described by separate linear parameterizations 
whose slopes differ by a factor of approximately 1.3\cite{NA4902}.}
\label{Entropy Plot}
\end{figure}

Following Ref.\cite{NA4902}, the entropy densities for each beam energy, 
here approximated as $\langle \pi \rangle / \langle N_{part} \rangle$, with $\langle \pi \rangle = 1.5 (\langle \pi^{-} 
\rangle + \langle \pi^{+} \rangle)$ to account for the neutral pions, are plotted in 
Fig.~\ref{Entropy Plot} as a function of the Fermi energy variable, $F \equiv 
{(\sqrt{s_{NN}}-2m_{N})^{3/4} \over \sqrt{s_{NN}}^{1/4} }$.  E895 data are indicated by stars and the values are tabulated 
in Table \ref{Entropy Table}.  NA49 results for 40, 80 and 158 AGeV Pb+Pb collisions from the CERN SPS were obtained 
from Ref.\cite{NA4902}.  The number of participants from \cite{NA4902}, 
calculated using the Fritiof\cite{Ande93} model, are also listed in Table 
\ref{Entropy Table}.

The linear dependence of the entropy per participant nucleon as a function of F in the 
proton-proton(anti-proton) data is not evident for the full range of the heavy ion 
collision data.  At and below AGS energies, the heavy ion data lie below the p+p data, 
and appear to be approximately linear with F.  In Ref. \cite{NA4902}, the SPS results 
combined with RHIC results at much higher energies from the PHOBOS Collaboration show a 
linear trend with a slope that is approximately 1.3 times larger than at the lower 
energies.  There appears to be a transition in the region between the AGS and top SPS 
energies.  The third-order polynomial fit to the heavy ion data shown on 
Fig.~\ref{Entropy Plot} may indicate that a smooth trend with increasing F can accurately 
describe the excitation function without the need for a discontinuous jump, such as 
one might expect from a first-order phase transition.  Two more runs at the SPS with beam 
energies of 20 AGeV and 30 AGeV may be able to improve the resolution in this important 
transition region.

\begin{table} 
{\centering \begin{tabular}{cccccccc}
$\sqrt{s_{NN}}$ (GeV)& F (GeV$^{1/2}$) & $\langle N_{p} \rangle$ & 
$\langle \pi \rangle / \langle N_{p} \rangle$ \\  
\hline
2.630 & 0.644 & 364 & 0.2279 $\pm$ 0.0159 $^{+0.0124}_{-0.0165}$ \\
3.279 & 0.965 & 366 & 0.5012 $\pm$ 0.0111 $^{+0.0307}_{-0.0357}$ \\
3.838 & 1.190 & 365 & 0.7385 $\pm$ 0.0148 $^{+0.0341}_{-0.0362}$ \\
4.289 & 1.351 & 363 & 0.9364 $\pm$ 0.0170 $^{+0.0570}_{-0.0579}$ \\
\hline
8.830 & 2.452 & 349 & 2.6433 $\pm$ 0.0858 \\
12.280 & 3.099 & 349 & 3.9542 $\pm$ 0.0869 \\
17.260 & 3.821 & 362 & 5.2127 $\pm$ 0.1823 \\
\end{tabular} \par}
\caption{Tabulated mean number of pions per participant and Fermi energy variable for 
2, 4, 6, and 8 AGeV Au+Au collisions and 40, 80, and 158 AGeV Pb+Pb collisions from 
Ref.\cite{NA4902}. Statistical and systematic uncertainties are reported separately.} 
\label{Entropy Table}
\end{table}

\section{Summary}
Transverse mass and rapidity spectra of charged pions in 2-8 AGeV 0-5\% central Au+Au 
collisions have been measured by the E895 experiment.  The transverse mass spectra 
exhibit a low-pt enhancement which can be largely ascribed to the feed-down from late 
stage resonance decays.  Differences in the {\pip} and {\pim} spectra at low {\mtm} 
are not reproduced by RQMD, which does not include final state interactions such as 
the Coulomb interaction with the nuclear source.  The inverse slope parameters 
increase as a function of beam energy and appear to be charge independent at high 
{\mtm}.  The measured rapidity distributions show excellent forward-backward rapidity 
symmetry and are well described by a model which includes collective longitudinal 
flow, with a velocity that is common to both pions and protons emitted in these 
collisions.  The 4$\pi$ yields of pions, obtained by integrating the rapidity 
distributions, have been used to infer an initial state entropy which increases 
with beam energy and is consistent with a smooth non-linear trend as a function of F 
from the 2 AGeV Au+Au collisions to 158 AGeV Pb+Pb collisions at the SPS.


        This work was supported in part by the U.S.\ Department of Energy
under grants DE-FG02-87ER40331.A008, DE-FG02-89ER40531, DE-FG02-88ER40408,
DE-FG02-87ER40324, and contract DE-AC03-76SF00098; by the US National 
Science Foundation under Grants No.\ PHY-98-04672, PHY-9722653,
PHY-96-05207, PHY-9601271, and PHY-9225096; and by the
University of Auckland Research Committee, NZ/USA Cooperative Science 
Programme CSP 95/33.


\end{document}